\documentclass[aps,prb,reprint,amsmath,amssymb,floatfix,superscriptaddress,longbibliography]{revtex4-1}
\usepackage{graphicx}
\usepackage{enumerate}
\usepackage{appendix}
\usepackage{xcolor}
\usepackage{hyperref}
\usepackage[english]{babel}
\usepackage[markup=underlined]{changes}
\usepackage{xcolor}
\usepackage{mathtools}
\usepackage{bm}

\def \be{\begin{equation}}
\def \ee{\end{equation}}
\def \ba{\begin{array}}
\def \ea{\end{array}}
\def \bea{\begin{eqnarray}}
\def \eea{\end{eqnarray}}

\DeclareMathOperator{\std}{std}
\def \scalefig{0.98}

\definechangesauthor[color=red]{KN}
\definechangesauthor[color=blue]{CC}

\newcommand{\homedir}{.}

\begin{document}
\title{Reconstruction of Current Densities from Magnetic Images by Bayesian Inference} \author{Colin B. Clement}
\author{James P. Sethna}
%\affiliation{Department of Physics, Cornell University, Ithaca, NY 14853-2501, United States}
\author{Katja C. Nowack}
\affiliation{Laboratory of Atomic and Solid-State Physics, Cornell University, Ithaca, NY 14853, United States}

\date{\today}

\begin{abstract}
Electronic transport is at the heart of many phenomena in condensed matter physics and material science. Magnetic imaging is a non-invasive tool for detecting electric current in materials and devices. A two-dimensional current density can be reconstructed from an image of a single component of the magnetic field produced by the current. In this work, we approach the reconstruction problem in the framework of Bayesian inference, i.e. we solve for the most likely current density given an image obtained by a magnetic probe. To enforce a sensible current density priors are used to associate a cost with unphysical features such as pixel-to-pixel oscillations or current outside the device boundary. Beyond previous work, our approach does not require analytically tractable priors and therefore creates flexibility to use priors that have not been explored in the context of current reconstruction. Here, we implement several such priors that have desirable properties. A challenging aspect of imposing a prior is choosing the optimal strength. We describe an empirical way to determine the appropriate strength of the prior. We test our approach on  numerically generated examples. Our code is released in an open-source \texttt{python} package called \texttt{pysquid}.
\end{abstract}

\maketitle

\section{Introduction \label{Sec:Intro}}
Two-dimensional (2D) materials host a variety of electronic transport phenomena, many of which are associated with a non-trivial spatial structure of the current density in the material. A non-invasive way to image a 2D current density is to image the stray magnetic field produced by the current and infer the current density. To date, numerous magnetic imaging techniques have been used to image current densities including scanning SQUID~\cite{kalisky2013,nowack2013, vasyukov2013}, scanning Hall probe~\cite{dinner2007},
magneto-optics~\cite{pashitski1997}, nitrogen-vacancy (NV) centers in
diamond~\cite{chang2017, tetienne2017,ku2019,Thiel2016} and levitated Bose-Einstein condensates~\cite{Yang2017,Yang2020}.

Most magnetic imaging techniques probe a single magnetic field component in a plane at a fixed height above the sample. (A notable exception are NV centers in diamond, which can realize vector magnetic field imaging~\cite{BroadwayPRA2020}.) The relation between the current density and the measured magnetic image is defined through two convolutions: the Biot-Savart law relates the magnetic field to the current density and a convolution of the magnetic field with the point spread function (PSF) of the magnetic sensor relates the magnetic field to the output of the sensor. To obtain the current density from a magnetic image, the two convolutions have to be inverted. If the current density only varies in two dimensions, this inversion is in principle possible because current conservation relates the two in-plane components of the current. In practice, the inversion is a non-trivial task because the problem is ill-posed: experimental images contain noise and the finite scan height and PSF lead to a loss of spatial information. As a consequence noise dominates the reconstructed image at high spatial frequencies. There are many solutions that predict the data including noise perfectly, but most of these solutions are not physical. Therefore, a criterion for what constitutes a physically sensible solution is required. This criterion is imposed through different so-called priors or equivalently a regularization.

A detailed overview of existing methods for current reconstruction is given by Meltzer et al. \cite{meltzer2017}. The most intuitive method is to invert the convolutions directly in Fourier space~\cite{roth1989}, filtering high spatial frequencies that otherwise cause instability. However, the shape and cutoff frequencies of the applied filters limit the resolution of the reconstructed image in an uncontrolled way. Iterative conjugate gradient methods have also been employed~\cite{wijngaarden1998fast}, which are more stable to noise. However, the regularization is not well controlled. Feldmann~\cite{Feldmann2004} and Meltzer et al.~\cite{meltzer2017} have reported reconstruction procedures using Tikhonov regularization penalizing the Laplacian of the current dipole field (defined below), combined with a cross-validation-based choice of the regularization strength. Tikhonov regularization is an attractive method because it is analytically tractable in Fourier space, i.e. it corresponds to a filter in Fourier space that can be expressed in Fourier space (see below). This allows for computationally efficient solutions and theoretically motivated methods of choosing the regularization strength.

In the wider image reconstruction literature, a variety of priors have been developed that are not analytically tractable. To date, these have not been applied to current reconstruction. For example, a prior based on total variation of the signal penalizes oscillations in a solution, but not sharpness like Tikhonov regularization does. An additional complication when reconstructing current densities is that typically some current leaves and enters the imaged field of view. At the points along the image boundary where this happens, the current density is not conserved. This violates the assumption of conserved current, without which the problem is underconstrained. Meltzer et al. \cite{meltzer2017} have implemented mirror boundary conditions for accommodating currents which enter or leave the image. This method has again the advantage of being analytically tractable, but it is not faithful to the sample geometry unless the sample is mirror-symmetric.

Here we describe a procedure to reconstruct current density from magnetic images that enables the use of a wider class of priors than previous work, and can accommodate currents crossing the image boundaries. We formulate the reconstruction problem in a probabilistic framework suitable for Bayesian inference that utilizes a generative model of the data. This approach offers significant flexibility to make use of prior information about the current density, including the sample geometry. Previous methods have penalized the Laplacian of the current dipole field (equivalent to Tikhonov regularization), from which the current density is computed, or the components of the currents themselves. Here, we discuss a set of requirements such as rotational invariance which a prior ideally should obey. From this we show that a prior based on the Frobenius of the Hessian is better motivated than the commonly used Laplacian prior. The approach presented here could be extended to other quantitative imaging problems in physics, where a generative model connecting the image and the underlying physical quantities is known.

The priors we explicitly discuss include a Gaussian prior (equivalent to Tikhonov regularization) penalizing the Laplacian and the Frobenius of the Hessian, a total variation prior, which penalizes strong fluctuations in the current density, but does not necessarily blur sharp edges, and a finite support prior, which allows the user to specify areas in the field of view where no current flows. In addition, we show that we can accommodate currents crossing the image boundary through modeling current densities flowing outside the field of view based on the sample geometry. This reconstruction problem ultimately leads to a convex optimization problem which we solve using the Alternating Difference Method of Multipliers (ADMM)\cite{boyd2011distributed}.

An important aspect of implementing a prior is to choose the strength with which it is imposed. If the prior is too weak, any reconstruction method tends to overfit the data such that even the noise is reproduced. If the prior is too strong, the resulting reconstruction is typically too smooth or has other undesirable and unphysical features. For some regularizations, one can theoretically predict the optimal strength of a prior for a given model. Typically, these are based on considering a single functional of a relatively simple metric such as the standard deviation of the residuals. A similar approach is not available for most priors discussed here. Instead, we choose the strength of the prior through inspecting both the standard deviation of the residuals as well as their spatial structure in real and Fourier space. The approach is more empirical and as implemented here requires visual inspection. However, an advantage is that a sense of the accuracy of the model can be gained through inspection of the spatial structure of the residuals.

The paper is organized as follows. In Sec.~\ref{Sec:Problem}, we define the forward problem
and describe how we use Bayesian inference. In Sec.~\ref{Sec:Priors} we propose requirements a prior should fulfill, derive a new prior which satisfies them, and compare it to a previously studied prior. We explore Gaussian priors and introduce a total variation and a finite support prior. Finally (Sec.~\ref{sec:chooselambda}), we discuss how we choose the strength of a prior. In Sec.~\ref{Sec:ExternalModel}, we describe how we account for currents flowing outside the field of view. We benchmark our method using numerical results throughout. Details of the inversion algorithm, the implementation of finite support priors, and the external current models are presented in the appendix. The code is organized in a \texttt{python} module called \texttt{pysquid}, which is publicly available in a github repository\footnote{\url{https://github.com/colinclement/pysquid}}.

\section{Bayesian inference formulation of the current reconstruction problem \label{Sec:Problem}}

\subsection{Forward problem}

First, we describe the forward problem: calculating the magnetic image resulting from a known 2D current density. We assume that the magnetic sensor probes the component of the magnetic field perpendicular to and at a fixed height above the plane in which the current flows. We assume a sensitive area of the sensor that is small compared to the scan height so that the PSF can be ignored. However, the methods we present here can be generalized to include a PSF and this capability is included in the code. Furthermore, it is straightforward to apply our reconstruction procedure to other magnetic field components, and to allow for a finite thickness of the current carrying sheet,as long as the current density has only in-plane component and does not change along the vertical direction

We choose the coordinate system such that the 2D sheet lies in the $x-y$ plane at $z=0$. 
Then, the magnetic field produced by a 2D sheet current $\bm{j}$ at a position $\bm{r}=(x,y,z)$ above the sample is given by the Biot-Savart law.
Assuming no current sources and drains are present in the field of view, the $x-$ and $y-$ components of the current density $j_{x,y}$ obey current conservation: $\partial_x j_x + \partial_y j_y = 0$. We explicitly enforce current conservation by introducing a single scalar field $g(\mathbf{s})$ which only depends on two dimensions. Here $\bm{s} = (x',y',0)$ because the current density is constrained to the $x-y$ plane at $z=0$. From this scalar field, we calculate the current density as $\bm{j}(\bm{s})=\nabla\times g(\bm{s})\bm{\hat{z}}$ with $\bm{\hat{z}}$ the unit vector in the $z$-direction. 
  
The Biot-Savart law as a function of $g$ can be written as ~\cite{wijngaarden1996determination,meltzer2017}
\begin{equation}
    \bm{B}(\bm{r}) = \frac{1}{4\pi}\int_V \mathrm{d}\bm{s}~g(\bm{s}) 
    \frac{3\hat{\bm{n}}(\hat{\bm{z}}\cdot\hat{\bm{n}}) - \hat{\bm{z}}}
         {|\bm{r}-\bm{s}|^3},
    \label{eqn:bskernel}
\end{equation}
where $\bm{\hat{n}}=(\bm{r}-\bm{s})/|\bm{r}-\bm{s}|$. The kernel convoluted with $g$ in Eq.~\ref{eqn:bskernel} is the magnetic field of a point dipole oriented along the $z$- direction. $g$ can therefore be viewed as a decomposition of a 2D current density into circulating currents, which is why we refer to $g$ as the current dipole field. The magnetic sensor probes the $z$-component, $B_z(\bm{r})$, of the magnetic field in Eq.~\ref{eqn:bskernel}.

\subsection{Inverse Problem and Bayesian Inference \label{Sec:Bayesian}}
In the following, we consider a discrete rectangular grid in the $x-y$ plane with pixels centered at coordinates $\{\bm{r}_i\}$. We define a magnetic image vector $\bm{\phi}$ with values $\phi_i = B_z(\bm{r_i})$ and a current dipole field vector $\bm{g}$ with values $g_i = g(\bm{r_i})$ sampled on the same lateral coordinates.
The linearity of the Biot-Savart law allows us to write the relationship between $\bm{\phi}$ and $\bm{g}$ as $\bm{\phi} = M\bm{g}$ for some suitable linear operator $M\in\mathcal{R}^{N\times N}$ where $N$ is the number of pixels in the image $\bm{\phi}$.

Each value $g_i$ of the discrete current dipole field corresponds to the amplitude and orientation of a current that circulates along the boundary of the pixel centered at $\bm{r_i}$. From Eq.~\ref{eqn:bskernel} we can directly calculate the elements of $M$ as a function of height (see details in the appendix). 
While the matrix $M$ is impractical to store for any reasonable image size, the product $M\bm{g}$ can be efficiently computed using a Fast Fourier Transform (FFT). A PSF can be incorporated in $M$. This capability is included in the code, but a detailed discussion of this aspect will be presented elsewhere.

We assume that the experimental noise is independent and identically distributed for each pixel, so that our model for a measured magnetic image is
\begin{equation}
    \bm{\phi} = M\bm{g} + \bm{\eta}
    \label{eq:model}
\end{equation}
where $\eta_i$ is the noise for each pixel. We assume Gaussian noise with variance $\sigma^2$, i.e. the probability distribution of the noise is a normal distribution with zero mean and standard deviation $\sigma$: $p(\eta_i) \sim \mathcal{N}(0,\sigma^2)$. The noise causes the data to fluctuate around the model with characteristic distance $\sigma$. 
We can therefore define the likelihood $p(\bm{\phi}|\bm{g})$ of
measuring $\bm{\phi}$ given $\bm{g}$:
\begin{equation}
    p(\bm{\phi}|\bm{g}) = \frac{1}{(2\pi\sigma^2)^{N/2}}\exp\left(-\frac{1}{2\sigma^2}\|M\bm{g} - \bm{\phi}\|^2\right),
    \label{eqn:likelihood}
\end{equation}
where $N$ is the number of pixels in the image $\bm{\phi}$, and $\|\cdot\|^2$ is
the Euclidean $L_2$ norm. This likelihood $p(\bm{\phi}|\bm{g})$ is our model of
the data and will allow us to infer the current dipole field.

Our goal is to learn $\bm{g}$ after having measured $\bm{\phi}$. We therefore
need $p(\bm{g}|\bm{\phi})$ (called the posterior probability).
Bayes' Theorem tells us how to reverse the conditional probability in Eq.~\ref{eqn:likelihood}:
\begin{equation}
    p(\bm{g}|\bm{\phi}) = \frac{p(\bm{\phi}|\bm{g})p(\bm{g})}{p(\bm{\phi})}.
    \label{eqn:bayes}
\end{equation}
Here $p(\bm{g})$ is the prior probability, encoding a criterion for preferable
and physically sensible solutions. $p(\bm{\phi})$ is called the evidence and is useful for quantitatively justifying the selection of one model over another~\cite{mackay1992bayesian}.
The maximum likelihood solution, $\bm{g}^\star$, to the reconstruction problem is then the most likely $\bm{g}$ given $\bm{\phi}$:
\begin{equation}
    \bm{g}^\star = \mathrm{max}_{\bm{g}} ~p(\bm{g}|\bm{\phi}) =
    \mathrm{max}_{\bm{g}} ~p(\bm{\phi}|\bm{g})p(\bm{g}).
    \label{eqn:maximumlikelihood}
\end{equation}
At this stage of inference the evidence $p(\bm{\phi})$ can be left out as it
is independent of $\bm{g}$. Full treatment of Bayesian inference
including optimal model selection is described in detail by
Mackay~\cite{mackay1992bayesian}, but is beyond the scope of this work.

Next, we assume that the prior probability (`prior' for short) can be written as $p(\bm{g}) \propto \exp\left(
-\lambda^2 \ell(\bm{g})\right)$ for some non-negative cost function $\ell$
and real number $\lambda$ which controls the strength of the prior. Combined with Eqs.~\ref{eqn:maximumlikelihood} and \ref{eqn:likelihood} we find the maximum likelihood solution for a given $\lambda$ and $\ell$ as
\begin{equation}
    \bm{g}_\lambda = \text{min}_{\bm{g}} ~\frac{1}{2}\|M\bm{g} - 
        \bm{\phi}\|^2 + (\lambda\sigma)^2~\ell(\bm{g}),
    \label{eqn:currentinference}
\end{equation}
Inference of currents is now cast as minimizing the
negative log-posterior, or minimizing the distance between our model $M\bm{g}$
and the data $\bm{\phi}$, constrained by a cost function $\ell(\bm{g})$. 

It is instructive to demonstrate the necessity of a nontrivial prior
$p(\bm{g})$.  If we consider all solutions as equally preferable, i.e. $p(\bm{g}) \propto 1$, the solution of Eq.~\ref{eqn:currentinference} is given by
\begin{equation}
    \bm{g} = (M^TM)^{-1}M^T\bm{\phi}.
    \label{eqn:psuedoinverse}
\end{equation}
Here $(M^TM)^{-1}M^T$ is the pseudoinverse, i.e. the `closest' inverse to the singular $M$, which is calculated only from the eigenvectors of $M$ with non-zero eigenvalue. $M$ has at least one zero eigenvalue, as adding any constant to $\bm{g}$ does not change
$\bm{\phi}$. The pseudoinverse ignores this symmetry, but since
the Biot-Savart law is long-range, $M$ has in addition exponentially small eigenvalues corresponding to high spatial frequencies. The noise has support in all frequencies. As a consequence, the solution to  Eq.~\ref{eqn:psuedoinverse} is highly unstable as the pseudoinverse amplifies any amount of noise. More specifically, the pseudoinverse can yield solutions which fit the data and noise perfectly. There is a huge space of solutions $\bm{g}$ that overfit the data like this, and most of them oscillate rapidly throughout the image. The role of a non-trivial prior $p(\bm{g})$ is to restrict this space by using physical arguments to specify which solutions are more likely.

\section{Constructing Physically Motivated Priors \label{Sec:Priors}}
\subsection{Gaussian Priors}
A common choice for a prior is a Gaussian with a cost function that depends on a linear transform of $\bm{g}$ :
\begin{equation}
    p(\bm{g}) \propto e^{-\lambda^2 \ell(\bm{g})}=\exp\left(-\lambda^2 \|\Gamma \bm{g}\|^2\right),
    \label{eqn:gaussianprior}
\end{equation}
where $\Gamma$ is a linear operator. Gaussian priors are the conjugate prior to a Gaussian likelihood. This allows to write the explicit solution to Eq.~\ref{eqn:currentinference}:
\begin{equation}
    \bm{g}_\lambda = \left(M^TM + (\sigma\lambda)^2 \Gamma^T\Gamma\right)^{-1}M^T\bm{\phi}.
    \label{eqn:tikhonov}
\end{equation}
In this form, we can see that the role of $\Gamma$ is to overwrite the exponentially small eigenvalues of $M^TM$, regularizing the instability of the pseudoinverse in eqn.~\ref{eqn:psuedoinverse}. The variance of the noise $\sigma$ sets the scale for the regularization strength $\lambda$. We will discuss in detail how to choose $\lambda$ in section \ref{sec:chooselambda}. Tikhonov regularization as discussed in refs.~\cite{Feldmann2004,meltzer2017} for current reconstruction as well as the optimal Wiener filter~\cite{press1989numerical} are equivalent to choosing a Gaussian prior with corresponding choices of $\Gamma$.

The simplest choice for $\Gamma$ is the identity $\Gamma=\mathbb{I}$. In this case, the prior favors a small-magnitude solution, which is not often physically motivated. If $\Gamma$ corresponds to derivatives, the prior prefers smooth solutions. The Laplacian $\Gamma = D_x^2 + D_y^2$, where $D_{x/y}^2$ are the second derivatives operator in the $x/y$-directions, is a common choice for image reconstruction problems and has been discussed in the context of current reconstruction~\cite{Feldmann2004,meltzer2017}. The Laplacian is translation invariant, prefers small accumulated curvature, and its solution given in Eq.~\ref{eqn:tikhonov} can be efficiently computed in Fourier space~\cite{Feldmann2004,meltzer2017}. In the following we refer to the corresponding cost function as the Gaussian Laplacian (GL) cost function $\ell_\mathrm{GL}(g)$. 

We can interpret $\ell_\mathrm{GL}(g)$ by using $\bm{j} = \nabla\times g\bm{\hat{z}} = \partial_y g \bm{\hat{x}} - \partial_x g\bm{\hat{y}}$ and writing $\ell_\mathrm{GL}(g)$ in the continuum limit:
\begin{align}
    \ell_\mathrm{GL}(g) &= - \lambda^2 \int \mathrm{d}^2\bm{r}~\left(\partial_x^2 g + \partial_y^2g\right)^2\nonumber\\
    &= -\lambda^2\int \mathrm{d}^2\bm{r}~|\nabla \times \bm{j}|^2.
    \label{eqn:laplacianprior}
\end{align}
The second line assumes that current only varies in the $x-y$ plane. We see that this cost function prefers solutions with small accumulated circulation of current. However, it is not clear why we should penalize only the circulation of current.

\begin{figure}[htp]
    \includegraphics[width=\scalefig \columnwidth]{\homedir/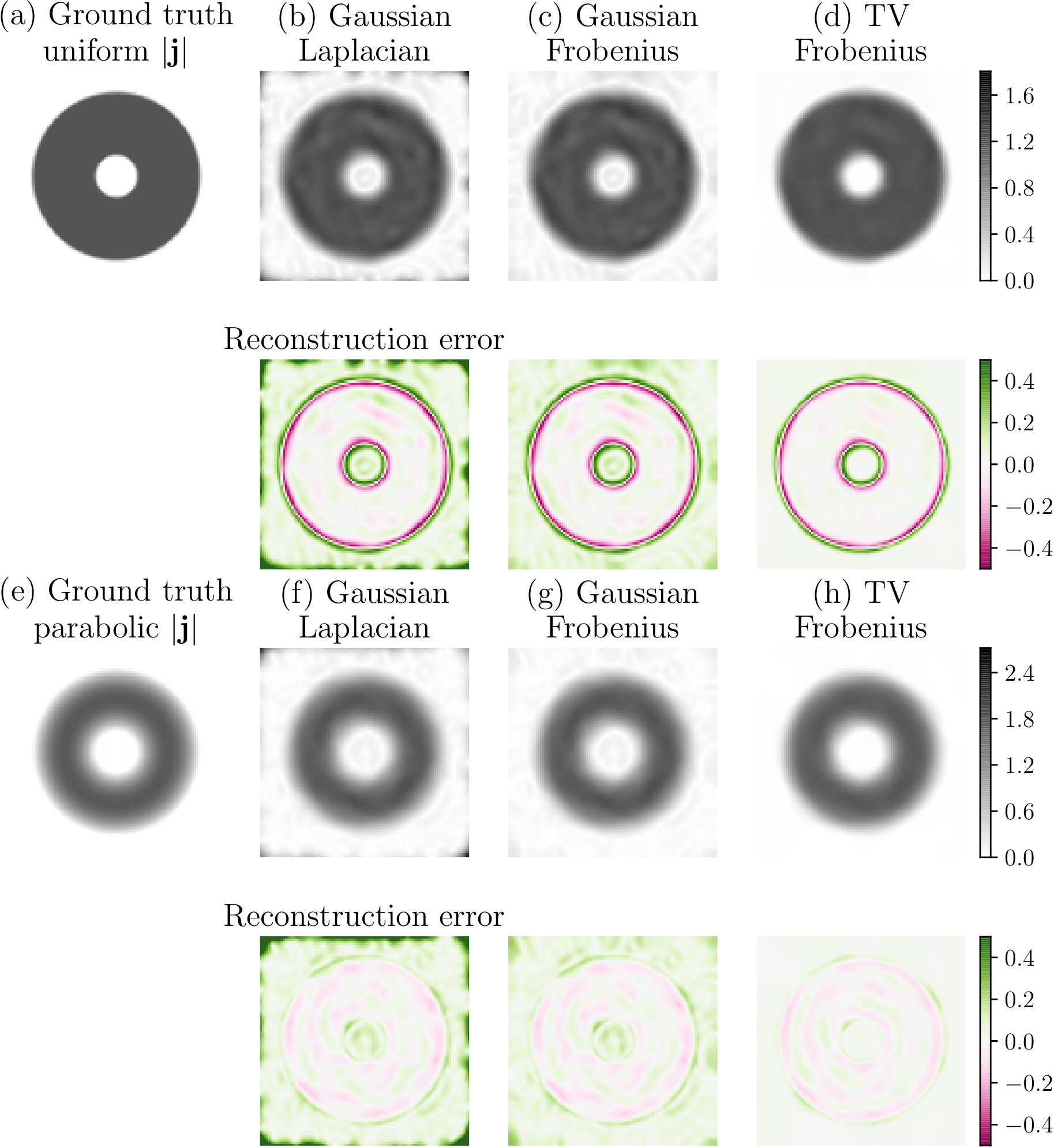}
    \caption{Ground truth current density $|\bm{j}|$, with uniform profile (a) and parabolic profile (e). The image size is 100$\times 100$ pixels. The magnetic images $\bm{\phi}$ was calculated at a height of 4 pixel widths. Noise of $\sigma=0.05$ was added. The reconstructed current density is shown for the uniform profile (b-d) and the parabolic profile (f-h) using a Gaussian prior penalizing the Laplacian (b,f), the Gaussian prior penalizing the Frobenius Hessian (e,g), and the total variation of the Frobenius Hessian (d,h). The regularization strength for each was chosen by the Bayesian discrepancy principle desribed in the main text. The data $\bm{\phi}$ was re-scaled to have unit peak-to-peak range, and $\lambda=1.4$ was used for the TVF prior reconstruction, and $\lambda=2$ was used for the GL and GF priors.}
    \label{fig:compare-priors}
\end{figure}

This leads us to more broadly explore choices for a Gaussian prior in the context of current reconstruction. For a prior that is chosen to suppress unphysical fluctuations in the reconstructed current density, it is desirable that the cost function is a functional of $g$ with physically-motivated symmetries and properties:
\begin{enumerate}
    \item Invariance under current inversions $g\rightarrow -g$,
    \item Invariance under rotations and reflections,
    \item Equally penalizing all variations in currents, i.e. first derivatives of $\bm{j}$ and thus second derivatives of $g$.
\end{enumerate}
The cost function $\ell_\mathrm{GL}$ satisfies almost all of these; it is the integral of $(\nabla^2 g)^2$; due to the quadratic it is invariant under $g\rightarrow
-g$, and it is rotation and reflection invariant (see also below). However, following Eq.~\ref{eqn:laplacianprior} we see that not all possible variations of the current are penalized.

We now derive a different cost function that satisfies all above constraints. The combination of the symmetry under current inversion and only allowing second derivative of $g$ constrains the functional to containing products of two elements of the second derivative matrix or so-called Hessian, $H$. This means that the prior must depend on the elements of the Hessian matrix $H_{\alpha\beta} = \partial_\alpha\partial_\beta g$, where $\alpha\in\{x,y\}$ and $\beta\in\{x,y\}$. One can show that the only way to construct a rotation invariant functional from the Hessian is through contraction of the indices~\cite{cvitanovic2008group} (following Einstein summation conventions). There are only two ways to do this~\cite{cvitanovic2008group} for products of two elements of $H$: (1) $H_{\alpha\alpha}H_{\beta\beta} = (\mathrm{Tr} H)^2$ and (2) $H_{\alpha\beta}H_{\alpha\beta} = \mathrm{Tr} H^T H$, where $\mathrm{Tr}$ denotes the trace of a matrix. The former is the Laplacian; the latter is the square of the Frobenius norm of the Hessian. This leads us to a new Gaussian Frobenius (GF) prior and corresponding cost function $\ell_\mathrm{GF}(g)$, which satisfies our first two criterion by construction. Following a similar calculation to Eq.~\ref{eqn:laplacianprior},
\begin{eqnarray}
  \ell_\mathrm{GF}(g) &=- \lambda^2 \displaystyle\int \mathrm{d}^2\bm{r}~& H_{\alpha\beta}H_{\alpha\beta}\nonumber\\
    &= - \lambda^2 \displaystyle\int \mathrm{d}^2\bm{r}~&
        \Big[(\partial_x j_y)^2 + (\partial_y j_x)^2+\nonumber\\
    &     &(\partial_x j_x)^2 + (\partial_y j_y)^2\Big],
    \label{eqn:frobeniusprior}
\end{eqnarray}
we find that $\ell_\mathrm{GF}(g)$ penalizes all variations in the current density and therefore also satisfies our third criteria.

We analyze the performance of $\ell_\mathrm{GF}(g)$ as a prior in Fig.~\ref{fig:compare-priors}. We study two annuli as synthetic data examples that realize different profiles of the current density across their width: a uniform profile (Fig.~\ref{fig:compare-priors}(a)) and a parabolic profile going to zero at the edges (Fig.~\ref{fig:compare-priors}(e)). The corresponding magnetic images were calculated assuming an imaging height above the plane of 4 pixel widths. Noise with $\sigma=0.05$ relative to the peak value in the magnetic image was added. Fig.~\ref{fig:compare-priors} shows reconstructions using the GL prior on the uniform annulus data (Fig.~\ref{fig:compare-priors}(b)) and the parabolic annulus data (Fig.~\ref{fig:compare-priors}(f)). Both reconstructions show large magnitudes of spurious currents at the edges of the image - likely due to the Laplacian not penalizing variations in all current components. Figs.~\ref{fig:compare-priors} (c) and (d) show reconstructions using the GF prior, yielding improved edge reconstructions. The strength $\lambda$ of the prior was set as described in section~\ref{sec:chooselambda}. Figs.~\ref{fig:compare-priors} (d) and (h) show reconstructions with a total variation prior using the Frobenius norm prior discussed below in section~\ref{Sec:TVprior}. These last reconstructions are more smooth where current is truly zero and have errors largely concentrated at the edges of the annuli.

We solved the maximum likelihood problem of Eq.~\ref{eqn:maximumlikelihood} with the Gaussian Laplace prior in Eq.~\ref{eqn:laplacianprior} and the Gaussian Frobenius prior in Eq.~\ref{eqn:frobeniusprior} by iteratively solving the appropriate regularization pseudoinverse in Eq.~\ref{eqn:tikhonov}. The construction of appropriate $\Gamma$ operators using centered finite difference derivatives is discussed in sec.~\ref{sec:linearoperators} of the Appendix. The computational complexity of one iteration of the solution of Eq.~\ref{eqn:tikhonov} is $\mathcal{O}(N\log N)$ for $N$ pixels in the data $\bm{\phi}$ via an FFT. The iterative method scales the same way, but will take a number of steps which depends on the condition number of the operator (which depends on $\Gamma$, $\sigma$, and $\lambda$).

\subsection{Total Variation Priors \label{Sec:TVprior}}
\begin{figure}[htp]
    \includegraphics[width=0.9 \columnwidth]{\homedir/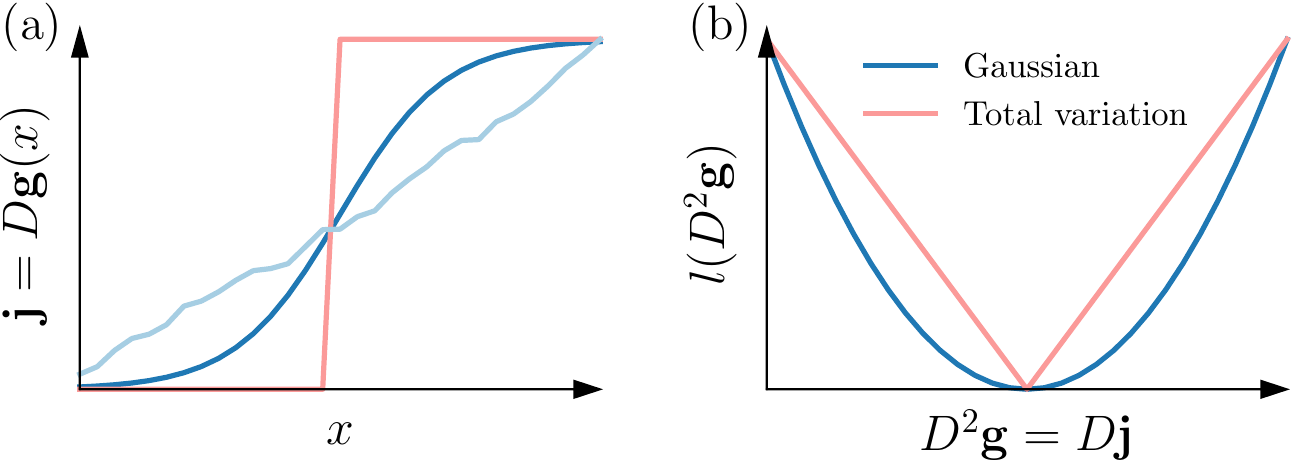}
    \caption{(a) Three different current $\bm{j}$ profiles which are equally probable under a total variation prior. (b) Illustration to compare Gaussian and total variation cost function. The Gaussian cost functions tolerates small variations in current, while the total variation cost function penalizes any non-zero amount.}
    \label{fig:tv}
\end{figure}

To our knowledge only analytically tractable, Gaussian priors have been considered in the context of current reconstruction. These are attractive because the resulting reconstruction problem can be solved using FFTs and there exist calculations for motivating the choice of regularization strength. However, Gaussian priors in particular suffer from ringing coming from sharp boundaries due to the Gibbs phenomenon~\cite{gottlieb1997gibbs}. For example, in Figs.~\ref{fig:compare-priors}(b) and (c), the Gaussian prior allows unnecessary variations of the current density inside the uniform annulus. Therefore, a prior which penalizes oscillations without penalizing sharp changes in the solution is desirable. In general image reconstruction, this is achieved by a so-called total variation (TV) prior, which depends on the sum of the absolute values of the derivative~\cite{vogel1998fast,osher2005iterative}. Since we want to penalize derivatives of $\bm{j}$, we need to penalize second derivatives of $g$. This leads us to a first possible choice for a total variation prior (ignoring momentarily the manifest violation of rotation invariance):
\begin{align}
    \ell_\mathrm{TV}(g) &=-\lambda^2 \int \mathrm{d}^2\bm{r}~
            |\partial^2_x g|+|\partial^2_y g|\nonumber\\
    &= -\lambda^2 \int \mathrm{d}^2\bm{r}~
            |\partial_x j_y|+|\partial_y j_x|.
    \label{eqn:originaltv}
\end{align}
We can gain some intuition about this original total variation cost function $\ell_\mathrm{TV}$ by considering a one dimensional profile. Fig.~\ref{fig:tv}(a) shows three hypothetical variations in current that are monotonic along one spatial direction. The TV prior considers all three equally probable regardless of their smoothness. The TV prior will suppress oscillations in the solution and, unlike Gaussian priors, remain agnostic to the sharpness of the transition. Fig.~\ref{fig:tv}(b)  shows both the GL and TV cost functions as functions of the second derivatives of $\bm{g}$. The GL cost function is more permissive of small variations of $\bm{j}$. In contrast, the absolute value of the TV causes any amount of variations $\bm{j}$ to be penalized. Therefore, $\ell_\mathrm{TV}$ prefers solutions of $\bm{g}$ with regions of constant $\bm{j}$ and allows sharp edges.

In Sec.~\ref{Sec:Priors} we argued that a cost function  should have several properties including rotational invariance. However, the TV cost function in Eq.~\ref{eqn:originaltv} is not rotation invariant. We identified that ideally a cost function should be a functional of the Frobenius norm of the Hessian of $g$: $\mathrm{Tr}H^TH$. This leads us to a TV Frobenius (TVF) cost function:
\begin{align}
    \ell_\mathrm{TVF}(g) &=-\lambda^2 \int \mathrm{d}^2\bm{r}
        \sqrt{H_{\alpha\beta}H_{\alpha\beta}}\nonumber\\
        &= -\lambda^2 \int \mathrm{d}^2\bm{r}
        \sqrt{(\partial_x^2g)^2 + (\partial_y g)^2 + 2(\partial_x\partial_yg)^2},
        \label{eqn:tvfrobenius}
\end{align}
where the square root of a sum of squares gives us a rotation invariant absolute value similar to the original TV cost function. 

Fig.~\ref{fig:compare-priors}(d) and (h) show the result of the TVF prior on the reconstructions of the uniform and parabolic current annuli. In both reconstructions, the background is uniformly zero as desired, since $\ell_\mathrm{TVF}$ penalizes any variation and zero current (a constant $g$ field) is effective at explaining the data. Likewise, in Fig.~\ref{fig:compare-priors}(d), the current density is more uniform in the interior of the annulus. In the case of a parabolic current density profile, the TVF prior (g) and Gaussian prior (h) perform  comparably. 

\subsection{Finite Support Prior\label{Sec:FiniteSupport}}
In many experiments lithographically defined devices are imaged for which the geometry is known in detail. If the field of view of the magnetic image contains regions of the device in which no current can flow, it is desirable to use this knowledge to improve the current reconstruction. Here, we describe and implement a prior that enforces zero current density in pre-defined regions of the image. Regions with zero current density correspond to regions of constant current dipole field $g$. 

In order to impose regions with constant $\bm{g}$, we define an image mask based on the device that identifies regions with zero current. The image mask $\bm{m}$ is assumed to be of the same shape as $\bm{\phi}$. It takes the value 0 in regions where current is unrestricted and therefore $\bm{g}$ can vary. It takes the value 1 where the current is zero and therefore $\bm{g}$ is constant. 

For each contiguous region in the mask with value 1, there is only one free parameter for the value of $g$ in that region. However, regions that are not connected can have different values. The total number of free values in $\bm{g}$ is then reduced to the number of contiguous regions of 0's in $\bm{m}$ plus the number of 1's in $\bm{m}$. We define $\tilde{\bm{g}}$ as a vector containing all free parameters in $\bm{g}$. There is a linear operator $F$ such that $\bm{g} = F \tilde{\bm{g}}$, where $F\in \mathbb{R}^{N\times P}$ for $N$ pixels in the image plane and $P$ free current dipole field parameters. $F$ is given by
\begin{equation}
    F_{jk} =
        \begin{cases*}
            1 & if free parameter $\tilde{g}_k$ sets $g_j$\\
            0 & else.
        \end{cases*}
\end{equation}
By replacing $\bm{g} \rightarrow F\tilde{\bm{g}}$ in Eq.~\ref{eqn:maximumlikelihood} we can impose regions with zero current as identified in the mask. We call this the `finite support' (FS) prior. It reduces the number of degrees of freedom and highly constrains the solution space. Since it can be implemented with a linear operator, it is straightforward to include it with both Gaussian priors and TVF prior discussed above. 

\begin{figure}[htp]
    \includegraphics[width=0.7\columnwidth]{\homedir/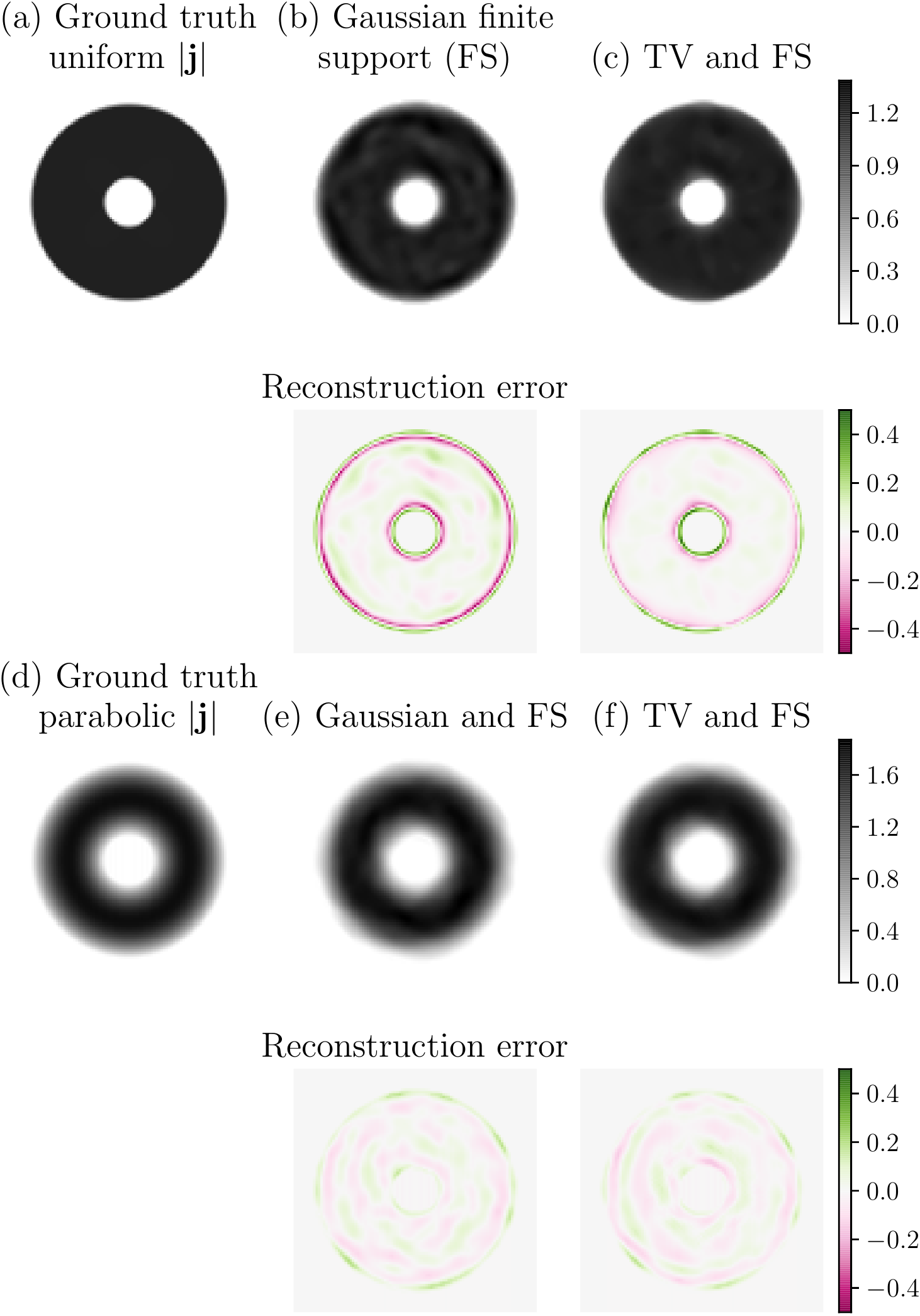}
    
    \caption{Ground truth current density $|\bm{j}|$, with uniform profile (a) and parabolic profile (d). The image size is 100$\times 100$ pixels. The magnetic image $\bm{\phi}$ was calculated at a height of 4 pixel widths and 5\% noise was added. The reconstructed current density is shown for the uniform profile in (b,c) and the parabolic profile in (e,f) using a Gaussian prior penalizing the Frobenius Hessian (b,e), and the total variation of the Frobenius Hessian (c,f). The regularization strength for each was chosen by the Bayesian discrepancy principle described in the main text. The data $\bm{\phi}$ was re-scaled to have unit peak-to-peak range, and $\lambda=1.4$ was used for the TVF prior reconstruction, and $\lambda=2.$ was used for the GF prior.}
    \label{fig:finitesupport}
\end{figure}

Fig.~\ref{fig:finitesupport} shows the result of adding the FS prior to the reconstruction of the numerical examples studied in Fig.~\ref{fig:compare-priors}. The values of $\bm{m}$ outside and inside of the annulus are set to 1 and the interior of the annulus to 0. One step of binary erosion is added in order to model experimental uncertainty in aligning the data to the lithography. Fig.~\ref{fig:finitesupport}(b) shows the reconstruction for the annulus with uniform current density profile using FS added to the GF prior: we see smooth edges and some remaining ringing inside the annulus. Fig.~\ref{fig:finitesupport}(c) shows another reconstruction using finite support added to the TVF prior: we see a very uniform interior current density and slightly sharper edges than in Fig.~\ref{fig:compare-priors}(d). For the annulus with a parabolic current density profile, Fig.~\ref{fig:finitesupport} shows that GF (e) and TVF (f) combined with FS yield very similar results. 

\subsection{Choosing the strength of the prior \label{sec:chooselambda}}

There are several methods for choosing the strength of the prior, including Bayesian evidence maximization~\cite{mackay1992bayesian}, the so-called $l$-curve method~\cite{hansen1992analysis}, cross-validation~\cite{golub1979generalized}, and the discrepancy principle~\cite{galatsanos1992methods}. Unfortunately, none of these methods works well for every prior. In fact, most methods require analytically tractable (Gaussian) priors. Since the TVF and FS priors are not analytically tractable, a more general method for setting the prior strength is needed. Here, we describe a modification of the discrepancy principle.

\begin{figure}[htp]
    %\centering
    \includegraphics[width=0.9 \columnwidth]{\homedir/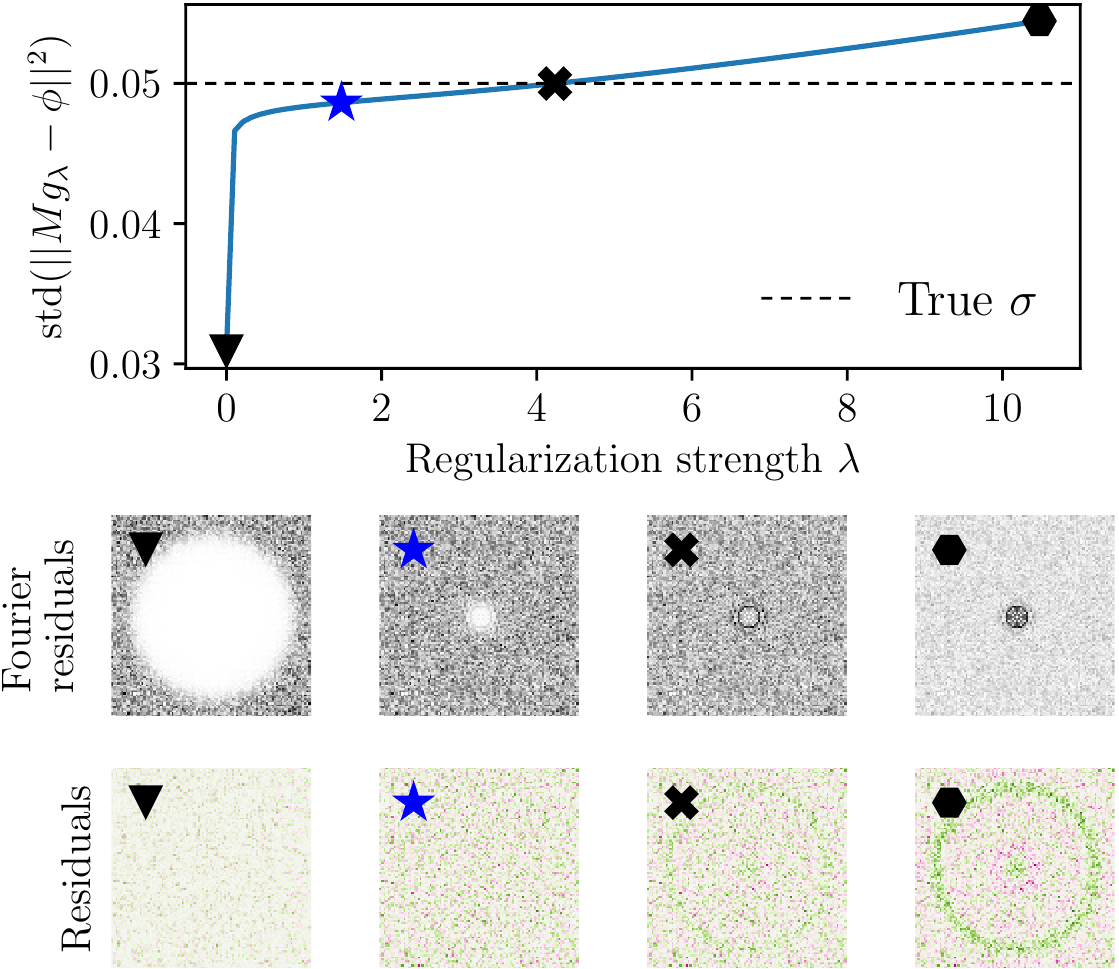}
    \caption{
    (Top) Example of standard deviation of the reconstruction error $\std\,(\|M\bm{g}_\lambda-\bm{\phi}\|^2)$ as a function of the strength of the prior $\lambda$ for the annulus with uniform profile and the GF prior from Fig.~\ref{fig:compare-priors} (c). Fourier amplitudes of the residuals (middle panels) and real-space residuals (bottom panels) are shown for the values of $\lambda$ as indicated by the symbols. The largest value of $\lambda$ for which the real space residuals have no spatial structure is denoted by a blue star. The black cross marks the value of $\lambda$ that satisfies the Bayesian discrepancy principle as discussed in the main text.  In order from left to right, we demonstrate the spatial properties of the reconstruction error for over-fitting of noise  ($\lambda\approx 0$), the result of the Bayesian discrepancy principal, under-fitting of the standard discrepancy principle, and finally an exaggeration of under-fitting.}
    \label{fig:reg-demo}
\end{figure}

Without a prior we can find a $\bm{g}$ for which the residuals $\bm{r} = M\bm{g}_\lambda-\bm{\phi}$ are arbitrarily small despite noise in the image. However, the corresponding $\bm{g}$ will have nonphysical properties such as rapid oscillations to reproduce the image including the experimental noise. The discrepancy principle is based on the observation that in an ideal situation the residuals should be given by the experimental noise. In its simplest form, the discrepancy principle therefore prescribes that the strength of the prior should be increased until the residuals have the same spectrum as the experimental noise. For Gaussian noise with variance $\sigma^2$, this implies adjusting $\lambda$ such that $\std\, (\|\bm{r}\|^2) = \sigma$. In Fig.~\ref{fig:reg-demo}~(top) we show the standard deviation of the residuals using the GF prior as a function of $\lambda$ for the annulus with uniform current density (see Fig.~\ref{fig:compare-priors}(a)). When inspecting images of the real-space residuals, we see that the residuals have spatial structure for the $\lambda$ that fulfills the discrepancy principle (black cross). The residuals should be independently distributed noise. This indicates that this simplest choice of the prior strength is too strong and produces a too smooth solution, which (partially) fails to reproduce sharp features in the image. Fig.~\ref{fig:reg-demo} (black hexagon) demonstrates the effect of an even stronger prior by showing the residuals for a larger $\lambda$. 

The discrepancy principle as described above can lead to over-smoothed solutions~\cite{galatsanos1992methods}. The cause is that the finite height in the Biot-Savart Kernel leads to some loss of spatial information. Therefore the true number of degrees of freedom that determine the image is actually lower than the number of pixels in the image. This can be taken into account by modifying the discrepancy principle. Let $\gamma<1$ be such that $\gamma N$ is the effective number of degrees of freedom, where $N$ is the number of pixels. Then the strength of the prior should fulfill $\std\,(\|M\bm{g}_\lambda-\bm{\phi}\|^2) = \gamma\sigma$. For Gaussian priors the $\gamma$ can in principle be estimated, but not for general priors. 

We therefore adopt a more empirical method to determine the optimum strength of the prior. We  find the regularization that satisfies the discrepancy principle, then reduce it until the residuals have minimal spatial structure in both real and Fourier space (as indicated by the blue star in fig.~\ref{fig:reg-demo}). Both are important: spatial structure in real space is indicative of a too strong prior, but for a too weak prior the residuals show no noticable spatial structure - only their distribution becomes more narrow. The FFT of the residuals shows the distribution of residuals across spatial frequencies. We see an increasingly pronounced spatial structure for weak strength of the prior. As the strength of the prior is reduced, noise corresponding to increasing spatial frequencies is fitted by the reconstruction. Therefore in Fourier space we can see intensity in the residuals missing up to a $k$-value that increases as the prior strength gets weaker. We refer to this modified discrepancy principle as Bayesian discrepancy principle, because it is guided by empirically identifying the most likely solution $\bm{g}_\lambda$.

In the example shown in Fig.~\ref{fig:reg-demo} we identify the value of $\lambda$ denoted by the blue star as the optimum prior strength. Some structure remains in the Fourier space residuals. In fact, there is no value of $\lambda$ at which there is truly no structure in the real-space and the Fourier transformed residuals. This is because our model including the prior is imperfect. Finally, if we choose $\lambda$ very small (indicated by the black triangle of fig.~\ref{fig:reg-demo}), we observe that the reconstruction fits the noise.

\section{Modeling currents outside the field of view \label{Sec:ExternalModel}}

\begin{figure}[htp]
    \includegraphics[width=0.85 \columnwidth]{\homedir/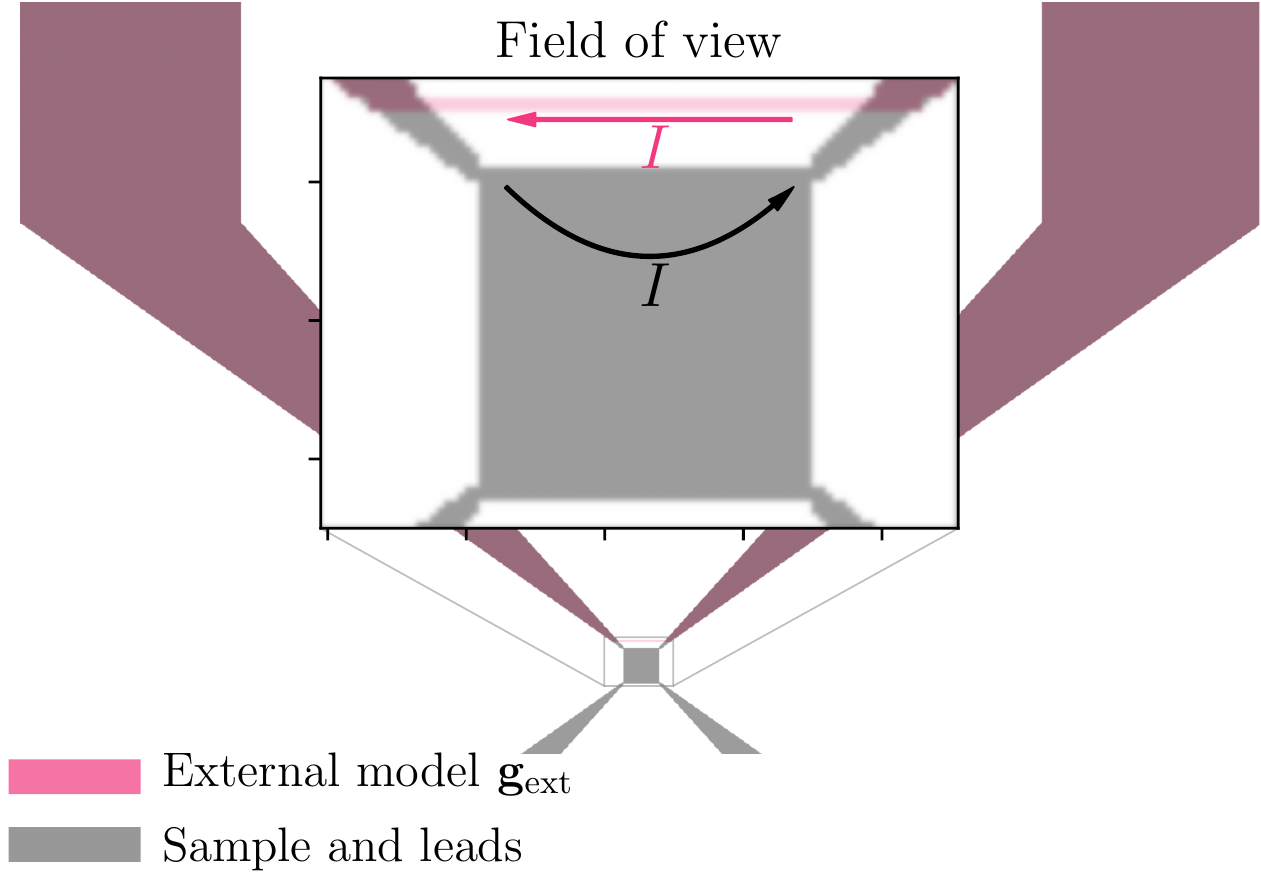}
    \caption{Schematic of a sample with van der Pauw geometry. Voltage is applied between the top two contacts, producing a current through the device with uniform resistivity. The sample is indicated in gray with the external model overlayed in pink. In order to remove currents crossing the field of view boundary, we calculate the field from the current density in the leads connected by a thin strip within the field of view. We subtract the resulting field $\bm{\phi}_\mathrm{ext}$ from the data $\bm{\phi}$. The currents associated with the external model flowing in the field of view are accounted for in the reconstruction.}
    \label{fig:external-schematic}
\end{figure}

\begin{figure*}[htp]
    \vspace{0.2cm}
    \includegraphics[width=0.9\textwidth]{\homedir/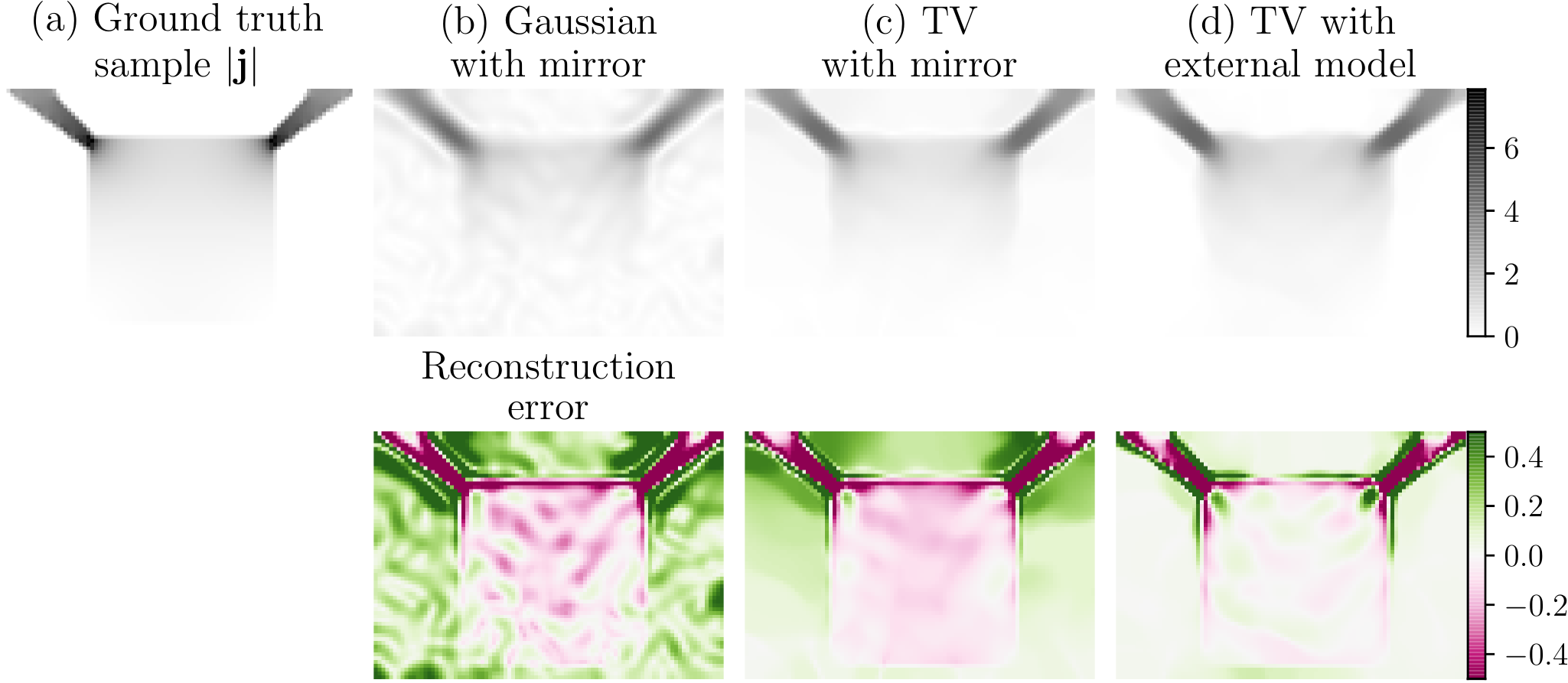}
    
    \caption{(a) Calculated current density for the sample shown in Fig.~\ref{fig:external-schematic} with uniform resistivity. Current enters and leaves at the top of the image. From this the magnetic field is calculated at a height of 4 pixel widths above the sample. Noise of $\sigma=0.05$ was added. The reconstructed current density is shown using (b) the GF prior with mirror-symmetric boundary conditions, (c) the TVF prior with mirror boundary conditions and (c) the TVF prior using an external model. The bottom row shows the residuals between the reconstructed current density and the ground truth in (a). The data $\bm{\phi}$ was re-scaled to have unit peak-to-peak range, and $\lambda=0.9$ was used for both TVF prior reconstructions and $\lambda=0.8$ was used for the GF prior.}
    \label{fig:compare-external}
\end{figure*}

A key assumption we have made so far is current conservation. We compute $\bm{j}$ from the current dipole field $\bm{g}$ as $\bm{j} = \nabla\times\bm{g}$. Therefore, our model describes currents that circulate within the field of view. However, when imaging a device, current enters and leaves the field of view in some parts of the image. One way to accommodate data in which current crosses the field of view is to assume mirror boundary conditions~\cite{meltzer2017}. Using mirror boundary conditions is analytically tractable, however in most cases it is not faithful to the sample geometry. In addition, it can be problematic, if the PSF of the magnetic sensor does not have mirror symmetry~\cite{nowack2013}.

Here, we include the option to model the current density outside the field of view. For many experiments the lithographic design of the imaged device is known. The field of view typically contains the active region of the device with metallic leads running outside of it. Assuming that the leads have a uniform resistivity, we build a loop which enters and leaves the field of view, canceling as much as possible the currents incident on the edges. As an example, we consider a device with a van der Pauw geometry (Fig.~\ref{fig:external-schematic}). The active region of interest in the device is the central square and we assume that the field of view is slightly larger than the square as shown. The leads of the device are long and widen which is typical to make it easy to attach wires to the device. We assume an externally applied current to flow from the top left to the top right lead. From this we calculate a current dipole field $\bm{g}$ for the full device and an image $\bm{\phi}$. In our example, we assume uniform resistivity in the square, but our method works for any current density in the device within the field of view. We build a model (shown in pink) that includes the leads and a segment within the field of view connecting them. We compute a corresponding current dipole field $\bm{g}_\mathrm{ext}$ and magnetic image $\bm{\phi}_\mathrm{ext}$ and subtract the latter from $\bm{\phi}$. The linearity of all equations guarantees that we will be only trying to recover the difference $\bm{g}-\bm{g}_\mathrm{ext}$, which should have conserved current in the field of view.

The external model requires extra variations in the optimal $\bm{g}$ which will be penalized by the cost function. Therefore, the cost function needs to be modified to be a function of $\bm{g}+\bm{g}_\mathrm{ext}$. We modify Eq.~\ref{eqn:maximumlikelihood} as follows:
\begin{align}
    \bm{g}_\lambda =~\bm{g}_\mathrm{ext} + 
        \text{min}_{\bm{g}}\Big[~\frac{1}{2}&\|M\bm{g} - (\bm{\phi}-\bm{\phi}_\mathrm{ext})\|^2 + \nonumber\\
        &(\lambda\sigma)^2~\ell(\bm{g}+\bm{g}_\mathrm{ext})\Big],
    \label{eqn:ext-inference}
\end{align}
where $\ell$ is any cost function. Appendix Sec.~\ref{sec:admm} explains how we accommodate these modifications for the TVF priors.

Using Eq.~\ref{eqn:ext-inference}, we can apply this method to include current leads for the device shown in Fig.~\ref{fig:external-schematic}. Fig.~\ref{fig:compare-external}(a) shows the current density in the field of view calculated for the entire device shown in Fig.~\ref{fig:external-schematic} with voltage applied to the top two contact pads. For this simulation, we used a tool that solves simple resistor networks included in the \texttt{pysquid} package. From the current density, we calculate $\bm{\phi}$ assuming a height above the plane of 4 pixel and add noise of magnitude $\sigma=0.05$. Fig.~\ref{fig:compare-external}(b) shows a reconstruction using the GF prior (see Eq.~\ref{eqn:frobeniusprior}), with mirror boundary conditions applied. The reconstruction shows some significant ringing in the current (see also Fig.~\ref{fig:compare-priors}). Figure~\ref{fig:compare-external}(c) shows the reconstruction using the TVF prior of eqn.~\ref{eqn:tvfrobenius} with mirror boundary conditions. We find significant current density outside the boundaries of the device, which is an artifact of the mirror boundary conditions since the device lacks mirror symmetry. Finally, Fig.~\ref{fig:compare-external}(d) shows the reconstruction using the external model of Eq.~\ref{eqn:ext-inference} with the TVF prior. This method significantly reduces the current density outside the boundaries of the device.

\section{Conclusions\label{Sec:Conclusion}}

The reconstruction of a current density from magnetic images is an increasingly important problem as more local magnetic probes are developed and applied to a variety of materials and devices. Optimal and flexible reconstruction methods are desirable to achieve optimal spatial resolution in the reconstructed image for a given magnetic probe. Even for a point-like magnetic probe, the height dependence of the Biot-Savart law introduces blurring. Therefore, current reconstruction requires strong regularization for stability. We followed the methods of the literature~\cite{wijngaarden1998fast,Feldmann2004}, defining the current dipole field $\bm{g}$ such that $\bm{j} = \nabla\times\bm{g}$, and defined the Biot-Savart kernel $M$ such that $\bm{\phi} = M\bm{g}$. Framing the reconstruction problem in terms of Bayesian inference, we defined the negative log-posterior in Eq.~\ref{eqn:currentinference}, the maximization of which provides a solution.

We discussed the importance of prior information, typically called regularization, and derived a new prior, the Frobenius of the Hessian, which improved the standard Gaussian prior. Real experimental data often contains sharp edges at which the current drops to zero (e.g. the  device boundaries) and areas of zero current (corresponding to areas where the device is absent). To improve the reconstruction of regions of constant current, we investigated a total variation prior and contrasted it with Gaussian priors which permit unnecessary oscillations. To leverage information about the device geometry we developed a finite support prior which can enforce where currents are guaranteed to be zero. Finally, we described a procedure to use a model for the currents outside of the field of view in order to accommodate the violation of current conservation in images of realistic samples.

With the total variation and finite support priors we moved beyond analytically tractable priors. We discussed how to choose the strength of the regularization using a Bayesian discrepancy principle. This method relies on well-defined criteria, but does require manual inspection. We argued that this is not a disadvantage compared to analytical tractable priors, for which an optimal prior strength can be explicitly estimated. Requiring manual inspection forces the user to engage with the data and the residuals. This can help to identify the quality of the model including the prior that is used. The flexibility to easily use different priors will then allow the user to evaluate how robust features in the reconstructed current density are for different priors and other parameters of the reconstruction.

In future work, we will develop methods to infer the PSF of a given imaging device. In principle, the PSF can be obtained from imaging a known source of magnetic field, e.g.  a magnetic dipole or a superconducting vortex. However, this is another ill-posed deconvolution problem in itself. The open-source \texttt{pysquid} code is already able to use a PSF into account. The package can be readily adapted to use additional priors. In particular, new opportunities arise from recent developments of machine learning such as deep priors~\cite{ulyanov2018deep} and random projectors~\cite{gupta2018random}. 

\begin{acknowledgments}
The authors thank Veit Elser, Matthew Bierbaum, Charlie van Loan and G. M. Ferguson for insightful discussions.
This work was primarily supported by the NSF (DMR-1719490). Support by the President's Council of Cornell Women (PCCW) Affinito/Stewart Program grants is acknowledged as well.  
\end{acknowledgments}

\section{Appendix}

\subsection{Numerically Implementing the Biot-Savart Kernel on a Discrete Grid}
In this section we derive the elements of the circulant matrix $M$ introduced in Eq.~\ref{eq:model}, which represents the convolution of the current dipole field with the Biot-Savart kernel~\cite{wijngaarden1996determination}. An image is defined through pixels arranged on a discrete, rectangular grid. We represent the sheet current density by rectangular pixels centered at $z=0$ below the position at which we detect the magnetic field. Rectangles of constant $g$ correspond to rectangular loop of current present only at the edges. We assume that the $z$-component of the magnetic field is detected. For a pixel centered at $\bm{s}_0 = (x_0, y_0, 0)$ with a value of  $g = 1$ the $z$-component of the magnetic field at $\bm{r}$ generated by this pixel is given by
\begin{equation}
    B^1_z(\mathbf{r},\bm{s}_0) = \frac{1}{4\pi}\int^{x_0+\frac{a}{2}}_{x_0-\frac{a}{2}}
                    \int^{y_0+\frac{b}{2}}_{y_0-\frac{b}{2}}
                    d\bm{s}                    \frac{3z^2-(\bm{r}-\bm{s})^2} {\left|\bm{r}-\bm{s}\right|^{5}},
    \label{eqn:onesquare}
\end{equation}
where $a$ and $b$ are the widths of the rectangle.

Following Ref.~\cite{wijngaarden1996determination}, we find the magnetic field due to a rectangle of constant $g$:
\begin{eqnarray}
    B^1_z(\mathbf{r}, \mathbf{s}_0) = \frac{1}{4\pi}\Big[
    &\mathcal{I}(x_0-x+\frac{a}{2}, y_0-y+\frac{a}{2},z)\nonumber\\
  -&\mathcal{I}(x_0-x+\frac{a}{2}, y_0-y-\frac{a}{2},z)\nonumber\\
  -&\mathcal{I}(x_0-x-\frac{a}{2}, y_0-y+\frac{a}{2},z)\nonumber\\
  +&\mathcal{I}(x_0-x-\frac{a}{2}, y_0-y-\frac{a}{2},z)\Big],
    \label{eqn:b1}                                             
\end{eqnarray}
where we defined 
\begin{equation}
    \mathcal{I}(x,y,z) =\frac{xy(2z^2+x^2+y^2)}{(z^2+x^2)(z^2+y^2)|\mathbf{x}|}.
\end{equation}

Next we write the magnetic field at position $\{x_k,y_l\}$ at height $z$ above a current distribution as a sum over the contributions from each individual pixel:
\begin{equation}
    B_z(x_k,y_l, z)=\sum_i\sum_j  B^1_z(x_k,y_l,z,x^\prime_i,y^\prime_j)g(x^\prime_i,y^\prime_j),
    \label{eqn:bconvolve}
\end{equation}
where $g_{ij}=g(x_i,y_j)$ for some set of pixel centers $\{x_i,y_j\}$. Since Eq.~\ref{eqn:b1} only depends on relative distances, we observe that Eq.~\ref{eqn:bconvolve} is a discrete convolution. This means that we can write our model more as $b=Mg$, where $b$ is the unraveled magnetic field image, $M$ is a circulant matrix representing the discrete convolution, and $g$ is our unraveled current dipole density image.

Because $M$ is circulant it is diagonalized by plane waves, and matrix-vector products like $Mg$ can be computed efficiently using Fourier transforms. A model for the PSF can in principle be included in the definition of $M$, as two discrete convolutions are a sequence of multiplications in Fourier space. 

\subsection{Defining Linear Operators to Numerically Implement Priors \label{sec:linearoperators}}

Here we discuss how we numerically implement the Gaussian Laplacian prior of eqn.~\ref{eqn:laplacianprior}, the Gaussian Frobenius prior of eqn.~\ref{eqn:frobeniusprior}, and the TV Frobenius prior of eqn.~\ref{eqn:tvfrobenius}. All priors contain partial derivatives $\partial_x^2$, $\partial_y^2$, and $\partial_x\partial_y$. To obtain discrete representations of these operators, we use finite-difference derivatives, encoded as a sparse matrix, where the interior of the image is computed using centered finite differences~\cite{press1989numerical}, and the edges use forward or backward finite differences, moving away from the edges. This way we can estimate the derivatives using only information that we have. As an example, we write the image $\bm{g}$ as a two-index matrix $g_{x,y}$, and derive $D^2_x$ explicitly. For a pixel not at the edge, and assuming that the distance between adjacent pixels is $\Delta x$,
\begin{equation}
    (D^2_x)_{x',y',x,y} = \frac{\delta_{x',x-1} - 2 \delta_{x',x} + \delta_{x',x+1}}
        {(2\Delta x)^2},
\end{equation}
whereas for example at the left edge,
\begin{equation}
    (D^2_x)_{0,y',x,y} = \frac{\delta_{0,x-2} - 2 \delta_{0,x-1} + \delta_{0,x}}
        {(2\Delta x)^2}.
\end{equation}
Here $\delta_{x,x'}$ denotes the Kronecker delta. We can similarly write discrete linear operators $D^2_x$, $D^2_y$ and the cross-derivative $D^2_{xy}$. Products $(D_x^2\bm{g})_i = \sum_i (D^2_x)_{i,j}~g_j$ are unraveled using appropriate $i=(x,y)$ and $j=(x,y)$. All implementations are included in the \texttt{pysquid} source code.

With these linear operators defined, we can write the discrete representations of the Gaussian Laplace priors given in Eq.~\ref{eqn:laplacianprior} as
\begin{equation}
    \ell_{GL}(\bm{g})= -\lambda^2 \|\Gamma \bm{g}\|^2,
\end{equation}
where $\Gamma = D^2_x + D^2_y$, and $\|\cdot\|^2 = \sum_i \cdot_i^2$. 

Next, we  find the operator $\Gamma$ representing the Frobenius of the Hessian, $\mathrm{Tr}H^TH$ where $H_{\alpha\beta}=\partial_\alpha\partial_\beta g$. While the Laplacian Gaussian integrand in Eq.~\ref{eqn:laplacianprior} is a square of a sum, the Frobenius prior is a sum of squares. In order to write this in terms of some operator $\Gamma$, we need to stack the operators
\begin{equation}
    \Gamma =
      \begin{pmatrix}
        D^2_x \\ D^2_y \\ \sqrt{2}D^2_{xy}.
      \end{pmatrix}
\end{equation}
Written this way, we can write the discrete generalization of the Gaussian Frobenius prior in Eq.~\ref{eqn:frobeniusprior} as
\begin{align}
    \ell_{GF}(\bm{g}) & = -\lambda^2 \|\Gamma \bm{g}\|^2
        =-\lambda^2 (\Gamma\bm{g})^T\Gamma\bm{g}\nonumber\\
        &=-\lambda^2 \Big[\sum_i (D^2_x g)_i^2 + \sum_i (D^2_y g)_i^2 +\nonumber\\
        &~~~~~~~~~~~\sum_i (D^2_{xy} g)_i^2\Big].
    \label{eqn:discretefrobenius}
\end{align}
Finally, the total variation prior can be expressed by modifying Eq.~\ref{eqn:discretefrobenius} by
taking a square root of the sum to obtain
\begin{align}
    \log p(\bm{g})\propto -\lambda^2 \Big[&\sum_i (D^2_x g)_i^2 + \sum_i (D^2_y g)_i^2 +\nonumber\\ 
    &\sum_i (D^2_{xy} g)_i^2\Big]^{1/2}.
\end{align}

\subsection{ADMM for Total Variation Deconvolution \label{sec:admm}}

Alternating Difference Method of Multipliers (ADMM)\cite{boyd2011distributed} is a convex optimization algorithm which solves
\begin{eqnarray}
    \min_{\bm{x},\bm{z}}&~&\mathcal{F}(\bm{x}) + \mathcal{G}(\bm{z})\nonumber\\
    \text{subject to}&~&A\bm{x} + B\bm{z} = \bm{c},
    \label{eqn:admm}
\end{eqnarray}
for some scalar functions $\mathcal{F}$ and $\mathcal{G}$, appropriately-sized matrices $A$ and $B$, and vector $\bm{c}$. The only requirement for ADMM to provably solve Eq.~\ref{eqn:admm} is that $\mathcal{F}$ and $\mathcal{G}$ be convex in their arguments.

We can cast our problem defined in Eqn.~\ref{eqn:currentinference} into the standard form of Eqn.~\ref{eqn:admm} by identifying $\bm{x} \equiv \bm{g}$, and by setting
\begin{eqnarray}
    \mathcal{F}(\bm{g}) &= \frac{1}{2}\|M\bm{g}-\bm{\phi}\|^2.
    \label{eqn:errorterm}
\end{eqnarray}
Then the function $\mathcal{G}$ is the regularization term set by $-\log p(\bm{z})$. Our isotropic total variation prior penalizes second derivatives of the $\bm{g}$-field. Identifying $A$ with the $x$- and $y$-derivative matrices $D^2_x$ and $D^2_y$ and $D^2_{xy}$,
\begin{equation}
    A = \left(\begin{matrix}
        D^2_x \\ D^2_y \\ D^2_{xy}
        \end{matrix}\right),
    \label{eqn:twoderiv}
\end{equation}
setting $B=-\mathbb{I}$, and $\bm{c} = 0$, the constraint $A\bm{g} + B\bm{z} = \bm{c}$ is equivalent to
\begin{equation}
    \left(\begin{matrix}
        D^2_x \\ D^2_y \\ D_{xy}
        \end{matrix}\right) \bm{g} = \bm{z} = 
    \left(\begin{matrix}
        \bm{z}_x \\ \bm{z}_y \\ \bm{z}_{xy}
        \end{matrix}\right),
\end{equation}
where $\bm{z}$ is twice as long as $\bm{g}$, containing both the horizontal and vertical second derivatives of $\bm{g}$.
The final piece is then the total variation of the Frobenius norm of the Hessian:
\begin{equation}
    \mathcal{G}(\bm{z})= \lambda^2 \sum_i \sqrt{z_{x,i}^2 + z_{y,i}^2 + 2 z_{xy,i}^2}.
    \label{eqn:itvfunction}
\end{equation}

We can modify ADMM to include finite support priors by replacing $\bm{g}\rightarrow F\tilde{\bm{g}}$, and optimizing $\tilde{\bm{g}}$ instead of $\bm{g}$:
\begin{eqnarray}
    \min_{\tilde{\bm{g}},\bm{z}}&~&\mathcal{F}(F\tilde{\bm{g}}) + (\bm{z})\nonumber\\
    \text{subject to}&~&AF\tilde{\bm{g}} = \bm{z},
    \label{eqn:finite-support-admm}
\end{eqnarray}
where $\mathcal{F}$ is still as in Eq.~\ref{eqn:errorterm}, $\mathcal{G}$ is as in Eq.~\ref{eqn:itvfunction}, and $A$ is as in Eq.~\ref{eqn:twoderiv}. Equivalently, we can modify the kernel matrix $M\rightarrow MF$ and the second-derivative matrix $A\rightarrow AF$. The latter is how it is implemented in the code.

To use an external model as in Eq.~\ref{eqn:ext-inference}, we modify ADMM by setting $\bm{c} = -A\bm{g}_\mathrm{ext}$ for the external model current dipole field in the field of view $\bm{g}_\mathrm{ext}$.

%\bibliography{reconstruction}

%

\end{document}